\documentclass[aps,pra,11pt,tightenlines,amsmath,amssymb,amsfonts,superscriptaddress,nofootinbib,showpacs,notitlepage]{revtex4-1}
\usepackage{IonFLRWpaper}
\usepackage{empheq}

\def\negspace{\!}

\def\lsub#1#2{{\vphantom{#1}}_{#2} \negspace {#1}}
\def\rsub#1#2{{#1} \negspace {\vphantom{#1}}_{#2}}

\def\bra#1{\left\langle {#1} \right\rvert}
\def\ket#1{\left\lvert {#1} \right\rangle}

\def\ketsub#1#2{\rsub {\ket{#1}} {#2}}
\def\brasub#1#2{\lsub {\bra{#1}} {#2}}

\def\outprodsubsub#1#2#3#4{\ketsub {#1}{#3} \brasub{#2}{#4}}

\def\moutprod#1#2{\outprodsubsub{#1}{#2}m m}

\newcommand\conft{\zeta} 	
\newcommand\wind{f}		

\newcommand\init{i}
\newcommand\final{f}
\newcommand\atom{\mathrm{A}}
\newcommand\laser{\mathrm{L}}

\renewcommand\op[1]{#1}	

\graphicspath{{graphics/}}

\begin{document}

\title{Simulating quantum effects of cosmological expansion using a static ion trap}

\author{Nicolas C. Menicucci}
\affiliation{Perimeter Institute for Theoretical Physics, Waterloo, Ontario, N2L 2Y5, Canada}

\author{S. Jay Olson}
\affiliation{Centre for Quantum Computer Technology, School of Mathematics and Physics, The University of Queensland, St Lucia, QLD, 4072, Australia}

\author{Gerard J. Milburn}
\affiliation{Centre for Quantum Computer Technology, School of Mathematics and Physics, The University of Queensland, St Lucia, QLD, 4072, Australia}

\begin{abstract}
We propose a new experimental testbed that uses ions in the collective ground state of a static trap for studying the analog of quantum-field effects in cosmological spacetimes, including the Gibbons-Hawking effect for a single detector in de~Sitter spacetime, as well as the possibility of modeling inflationary structure formation and the entanglement signature of de~Sitter spacetime.  To date, proposals for using trapped ions in analog gravity experiments have  simulated the effect of gravity on the field modes by directly manipulating the ions' motion.  In contrast, by associating laboratory time with conformal time in the simulated universe, we can encode the full effect of curvature in the modulation of the laser used to couple the ions' vibrational motion and electronic states.  This model simplifies the experimental requirements for modeling the analog of an expanding universe using trapped ions and enlarges the validity of the ion-trap analogy to a wide range of interesting cases.
\end{abstract}

\date{August 2, 2010}


\maketitle

\section{Introduction}
\label{sec:intro}

Curved-spacetime quantum field theory~(QFT) is an essential ingredient of our modern understanding of cosmology~\cite{weinberg2008} and has also given rise to remarkable insights and puzzles concerning the nature of Planck-scale physics~\cite{hawking1975}.  Unfortunately, most predictions of curved-spacetime QFT are extremely difficult to test directly, confined as we are to a low-curvature region of spacetime, while dimensional analysis conspires to make the most straightforward effects extremely tiny.  This situation has given rise to a number of proposals for testing \emph{analog} curved-spacetime QFT effects, where spacetime is replaced by a system of coupled atoms, and field modes are replaced by the collective normal modes of oscillation of the atoms---effectively a ``phonon field.''  The advantage is that while the observable effects are conceptually the same (analogous), the parameters of the experiment may be adjusted to result in a much stronger and more easily observed effect.  The most common analog curved-spacetime QFT proposals involve BECs, liquid helium, or ion traps as the analog spacetime~\cite{Schutzhold2009, barcelo2005}.

Here, we employ a general but often overlooked concept to extend the range of possible analog spacetime QFT experiments in ion traps, and we discuss several analog cosmology experiments that appear to be greatly simplified by the new picture, proposing one of them (simulation of the Gibbons-Hawking effect~\cite{Gibbons1977}) in detail.  The main idea, which has been proposed previously in the context of BEC simulations~\cite{Fedichev2003,Fedichev2004}, is to encode the effects of spacetime curvature into a modulation of the analog detectors, rather than in the motion of the atoms which comprise the analog spacetime.  We will show how this amounts to making a shift in the analogy away from $\text{\emph{laboratory time}} \leftrightarrow \text{\emph{detector proper time}}$ and towards $\text{\emph{laboratory time}} \leftrightarrow \text{\emph{conformal time}}$.  In general, such a shift in analogy can be made between the experimental lab time and any new time coordinate in the simulated spacetime, but in the case of Friedmann-Lema\^itre-Robertson-Walker~(FLRW) cosmologies in the conformal vacuum, we will see that this particular choice (laboratory time made analogous to conformal time) encodes the entire effect of the spacetime curvature and therefore enables us to propose analog curved-spacetime QFT experiments in which the analog spacetime is fixed (resembling flat Minkowski space), thus making our proposals far more accessible experimentally.

Building on efforts to implement elementary quantum gates, ion trapping now offers some of the most exquisite demonstrations of quantum control in modern physics.  The ability to couple the internal electronic state of one or more trapped ions to the collective vibrational degrees of freedom enable the ability to produce a large class of pure quantum states~\cite{Leibfried2003,Monroe1995,James1998}, including Fock states, squeezed states, and cat states, as well as quite complex information processing tasks~\cite{Leibfried2003a,Schmidt-Kaler2003}.  We use the control of the trapping potential afforded by ion traps, together with the ability to reach quantum limited motion, to propose the construction of new ion-trap analogs of the single-detector response in FLRW cosmology (including the Gibbons-Hawking effect~\cite{Gibbons1977}), along with several proposals for ways to apply the model to situations involving multiple detectors, including inflationary de Sitter cosmology~\cite{Bardeen1983, Bassett2006} and the entanglement signature of de~Sitter spacetime~\cite{VerSteeg2009}.  We discuss the most important experimental parameters governing the simulation and the regime in which the analog experiment is likely to conform closely to the simulated cosmology. Highly efficient quantum measurements, based on fluorescent shelving~\cite{Leibfried2003}, provide a practical means to test our predictions.  The main advantage of trapped-ion analog gravity experiments over other proposals, such as BECs~\cite{Fedichev2003,Fedichev2004}, lies in the exquisite quantum control over state preparation, interaction, and readout.  In particular, the fact that each ion is equipped with its own highly efficient and tunable motional detector (through laser coupling~\cite{James1998}), which counts individual phonons, means that ion traps have a particular advantage in efficient quantum control of the detector interaction and readout~\cite{Schutzhold2009}.  We compare our proposal to previous work~\cite{Alsing2005,Schutzhold2007} that suggests using ion traps with a time-dependent trapping potential to simulate cosmological particle creation.

We organize the paper in the following way:  Section~\ref{sec:confinv} describes in more detail the conceptual shift from the standard ``field picture'' of analog QFT to the ``detector picture,'' which we use extensively, and reviews the relevant properties of FLRW cosmology that make it possible.  In Section~\ref{sec:cosmologyions}, we review the fundamental features of ion-trap simulations of cosmology, which are essential for describing our proposals.  Section~\ref{sec:apps} is devoted to the description of our experimental proposals.  Our concluding remarks are contained in Section~\ref{sec:conc}.  Unless otherwise specified, we use natural units, in which~$\hbar = c = 1$.

\section{Conformal Invariance: Field Picture and Detector Picture}
\label{sec:confinv}

To build an experimental analogy to quantum field effects in spacetime, we will make use of two basic ingredients: (1)~a coupled, discrete system of oscillators, where deviation of a given oscillator from its equilibrium position $\op{q}_{n}(t)$ plays the role of the field $\op{\phi}(\vec x, t)$, and (2)~one or more detectors, which can be coupled to the analog field in a way which models a given type of particle detector in spacetime.  In the case of the ion traps we discuss in the next section, for example, the electronic state of a specific atom is coupled to the deviation from its equilibrium position.  

Before we discuss the analog experiment, however, we recall a few basic properties of standard Unruh-DeWitt detectors in FLRW spacetime~\cite{Birrell1982}.  Recall that the usual coupling of the field to detector is expressed via the following interaction Hamiltonian (which, importantly, assumes that the cosmic time variable~$t$ equals the proper time~$\tau$ for the detector):
\begin{align}
\label{eq:Hint}
	H_{\text{int}}(t) = f(t) \op{m}(t) \op{\phi}(\vec x,t)\,,
\end{align}
where all operators are in the interaction picture, $\op{m}(t)$~is the detector monopole moment, and $f(t)$~is a ``window function'' modeling a time-dependent coupling of the field to the detector.  In the case of a single detector with energy gap~$E$ at a fixed location~$\vec x$, this leads to a detector response function given by
\begin{align}
\label{eq:Ageneric}
A =  \int_{- \infty}^{\infty} dt \int_{ - \infty}^{\infty} dt' f(t) f(t') e^{i E (t - t')} \bar D^{+}(t, t')\,.
\end{align}
This is the probability that the detector, starting in its ground state~$\ket g$, will become excited into its higher-energy state~$\ket e$ after interacting with the field through application of the Hamiltonian~$H_{\text{int}}(t)$ over a very long time---i.e.,~at least long enough to cover the (nontrivial) support of~$f(t)$.  This formula assumes the interaction is weak enough for time-dependent perturbation theory to be used.  Here, $\bar D^{+}(t, t')$~is the two-point function given by
\begin{eqnarray}
\bar D^{+}(t, t') = \langle \op{\phi}(\vec x, t) \op{\phi}(\vec x, t') \rangle
\end{eqnarray}
for a given state of the field (recall that the detector is fixed at spatial position~$\vec x$), and the bar is used with foresight to indicate that the time coordinate being used is~$t$ [Cf.~Eq.~\eqref{eq:Dconformal}, below].

An analog experiment is one which reproduces the essential features of the response function in curved spacetime for an analogous detector coupling and state of the oscillators (typically with the ground state of the collective normal modes of the oscillators being analogous to a vacuum state of the spacetime field).  What we would like to emphasize here is that this can be achieved in multiple ways: by manipulating the collection of oscillators themselves or by manipulating the detector.  When the spacetime involved is conformal to Minkowski spacetime, as in FLRW cosmology, the \emph{entire effect of curvature} can be encoded completely within either the oscillators or the detector, leading to two distinct ``pictures'' for analog experiments.

\subsection{The Field Picture}

In the field picture, the relevant features of the analog spacetime are encoded in the oscillators and thus directly in the modes of the analog field and the resulting two-point function.  Thus, in the response function, the analogy is contained in the two-point function directly, i.e.,~$\langle \op{\phi}(\vec x, t) \op{\phi}(\vec x, t') \rangle \leftrightarrow \langle \op{\phi}_{n}(t) \op{\phi}_{n}(t') \rangle$ (where $\op{\phi}_{n}(t)$~is an appropriately dimensionless version of~$\op{q}_{n}(t)$).  The analogy is completed by physically moving the oscillators---an expanding array of ions simulating an expanding spacetime, for example~\cite{Alsing2005,Schutzhold2007}.  The important feature of this picture is that $t$~labels both the proper time of the analog detector in the lab and the proper time of the detector being simulated.

The field picture is thus immediate conceptually, and for this reason it has been the ``default'' picture for proposed analog cosmological QFT experiments in the literature.  We emphasize, however, that it is not the only picture for cosmological models and that adequate control of the analog spacetime (the oscillators) often presents a set of significant experimental challenges.

\subsection{The Detector Picture}

When a given spacetime is conformal to Minkowski spacetime, and when the fields considered are massless and conformally coupled, we can switch to the \emph{detector picture}, in which the array of oscillators is held fixed (in analogy to Minkowski spacetime), but the detector is modified in a time-dependent fashion to create the same analog effect as a curved spacetime.  The important conceptual feature of this picture is that the laboratory proper time describing the evolution of the analog detector is \emph{not} analogous to the proper time of the simulated spacetime detector.  Instead, the experimental proper time (i.e.,~the time measured by the clock on the wall in the laboratory) is analogous to the \emph{conformal time} of the simulated spacetime and detector.  This is accomplished by introducing a time-dependent energy gap in the analog detector, which simulates an ordinary, constant-energy gap of a detector in a curved cosmological spacetime.  The payoff for the conceptual shift and the modulation of the detector is a greatly simplified treatment of the analog spacetime---the array of oscillators is held fixed.

To see how this works, we recall the transformation properties of fields in spacetime (see Ref.~\cite{Birrell1982} for more details).  For simplicity, we work with a scalar field.  Recall that a spacetime (with metric $g_{\mu \nu}$) conformal to Minkowski spacetime (wih metric~$h_{\mu \nu}$) can be coordinatized such that $g_{\mu \nu}(\vec x, \conft) =  \Omega(\vec x, \conft)^{2} h_{\mu \nu}$.  (In this section $x = (\vec x, \conft)$, with $\conft$~playing the role of a time coordinate.)  In these coordinates, the two-point function in the conformal vacuum satisfies
\begin{align}
\label{eq:Dconfgeneral}
		D^{+}[\vec x(\conft), \conft; \vec x'(\conft'), \conft'] = \Omega(x)^{(2 - n)/2} {D}^{+}_{\text{M}}[\vec x(\conft), \conft; \vec x'(\conft'), \conft'] \Omega(x')^{(2 - n)/2}\,,
\end{align}
where ${D}^{+}_{\text{M}}[\vec x(\conft), \conft; \vec x'(\conft'), \conft']$~is the standard Minkowski two-point function for a scalar field along the trajectory $\vec x(\conft)$, $n$~is the number of spacetime dimensions, and we omit the bar over~$D^{+}$ because these are functions of~$\conft$ instead of~$t$ [Cf.~Eq.~\eqref{eq:Ageneric} above and Eq.~\eqref{eq:Dconformal} below].  The variable~$\conft$ is thus interpreted as the conformal time of an FLRW universe on the left-hand side of the above equation and the Minkowski time on the right-hand side.

In FLRW cosmology with no spatial curvature, the metric in comoving coordinates takes the form
\begin{align}
\label{eq:FLRWmetriccosmic}
	ds^{2} = dt^{2} - a(t)^{2} \sum_{i= 1}^{n - 1} (dx^{i})^{2}\,,
\end{align}
where $t$~is cosmic time and also the proper time for a comoving observer (i.e.,~an observer on an inertial trajectory of constant~$\vec x$).  Transforming to conformal coordinates by substitution of the conformal time~$\conft$, which satisfies
\begin{align}
\label{eq:conft}
	d \conft = \frac {dt} {a(t)}\,,
\end{align}
results in a metric of the following form:
\begin{align}
\label{eq:FLRWmetricconformal}
	ds^{2} = \Omega(\conft)^{2} \left[ d \conft^{2} - \sum_{i = 1}^{n - 1} (dx^{i})^{2} \right] \,,
\end{align}
where
\begin{align}
\label{eq:Omegadef}
	\Omega(\conft) = a[t(\conft)]\,.
\end{align}
By using Eq.~\eqref{eq:Dconfgeneral}, the two-point function on the trajectory of a comoving detector at a fixed location~$\vec x$ can be expressed as
\begin{align}
\label{eq:Dconformal}
	\bar D^{+}_{\text{FLRW}}(t,t') = D^{+}_{\text{FLRW}}(\conft, \conft') = \Omega(\conft)^{(2-n)/2} \Omega(\conft')^{(2-n)/2} D^{+}_{\text{M}}(\conft, \conft')\,,
\end{align}
where $D^{+}(\conft, \conft')$~is short for~$D^{+}[\vec x(\conft), \conft; \vec x'(\conft'), \conft']$ for both the FLRW and Minkowski Green's functions as functions of~$\conft$, and we explicitly show the connection to the Green's function~$\bar D^{+}(t,t')$ used in Eq.~\eqref{eq:Ageneric}.  The factor of~$\Omega(\conft)^{(2-n)/2} \Omega(\conft')^{(2-n)/2}$ may be absorbed into the definition of the detector window functions, and thus an analog Minkowski two-point function is equally good for describing FLRW cosmology, provided we can re-interpret the experimenter's ``clock time'' as the conformal time $\conft$.  For this interpretation to work, however, the detector must be designed to evolve with respect to the simulated proper time $t$ as $e^{- i E t}$ in order to model a comoving detector, rather than evolving in the simulated conformal time $\conft$ as $e^{- i E \conft}$.

The appropriate modification of the detector requires that we simulate time evolution in~$t$ with a \emph{$t$-independent} Hamiltonian $H_{0}$ by means of a \emph{$\conft$-dependent} Hamiltonian~$H_{0}(\conft)$.  This time-dependent Hamiltonian will be the physical Hamiltonian acting on an individual oscillator in the lab~\cite{Olson2010,Fedichev2003,Fedichev2004}.  To find~$H_{0}(\conft)$, we simply write the desired Schr\"odinger equation in~$t$ and transform to~$\conft$.  Thus, a time-independent~$H_{0}$ in the equation\begin{align}
\label{eq:Schrodingert}
	i \frac{d}{d t} \Psi &= H_{0} \Psi
\intertext{can be re-expressed as}
\label{eq:Schrodingerconft}
	i \frac{d}{d \conft} \Psi &= H_{0}(\conft) \Psi\,,
\end{align}
where~$H_{0}(\conft) = \frac{d t}{d \conft} H_{0}$.  Simulating a detector with fixed energy gap~$E$ in~$t$, given by~$H_{0}$, requires a time-dependent ``laboratory Hamiltonian''~$H_{0}(\conft)$, which corresponds to a simple time-dependent scaling of the energy gap in laboratory time.  For the case of a FLRW cosmology simulation, this is given by
\begin{align}
\label{eq:H0conft}
	H_{0}(\conft) = \Omega(\conft) H_{0} = a[t(\conft)] H_{0}\,.
\end{align}

Observe how these ingredients come together to describe the response of a comoving detector at a fixed location~$\vec x$ in an FLRW universe:
\begin{align}
\label{eq:AFLRW}
	A_{\text{FLRW}} &=  \int_{- \infty}^{\infty} dt \int_{ - \infty}^{\infty} dt' f(t) f(t') e^{i E (t - t')} \bar D^{+}_{\text{FLRW}}(t, t') \nonumber \\
	&= \int_{\conft(t=-\infty)}^{\conft(t=\infty)} d\conft \int_{\conft(t=-\infty)}^{\conft(t=\infty)} d\conft' \frac{d t}{d \conft}  \frac{d t'}{d \conft'} f[t(\conft)] f[t'(\conft')] e^{i E [t(\conft) - t'(\conft')]} \nonumber \\
	&\qquad \qquad \times \Omega(\conft)^{(2-n)/2} \Omega(\conft')^{(2-n)/2} D^{+}_{\text{M}}(\conft, \conft')  \nonumber \\
	&= \int_{\conft(t=-\infty)}^{\conft(t=\infty)} d\conft \int_{\conft(t=-\infty)}^{\conft(t=\infty)} d\conft' F(\conft) F(\conft') e^{i E [t(\conft) - t'(\conft')]} D^{+}_{\text{M}}(\conft, \conft')\,,
\end{align}
where
\begin{align}
\label{eq:windowfunc}
	F(\conft) &= \frac{d t}{d \conft} \Omega(\conft)^{(2-n)/2} f[t(\conft)] = a[t(\conft)]^{(4-n)/2} f[t(\conft)]
\end{align}
is the window function in $\conft$ (conformal time, also Minkowski time) necessary to simulate the window function~$f(t)$ in cosmic time.  Note the non-standard exponential factor in the last line, coming from the $\conft$-dependent energy gap.  This connection between the field picture and the detector picture is illustrated schematically in Figure~\ref{fig:fielddetectpics}.

\def\wavewidth{0.14\columnwidth}
\def\levelwidth{0.09\columnwidth}

\begin{figure}
\begin{center}
\begin{tabular}{@{}c@{\;}c@{\;}c@{\qquad}c@{\qquad}c@{\;}c@{\;}c@{}}%
\parbox[c]{\wavewidth}{\includegraphics[width= \wavewidth]{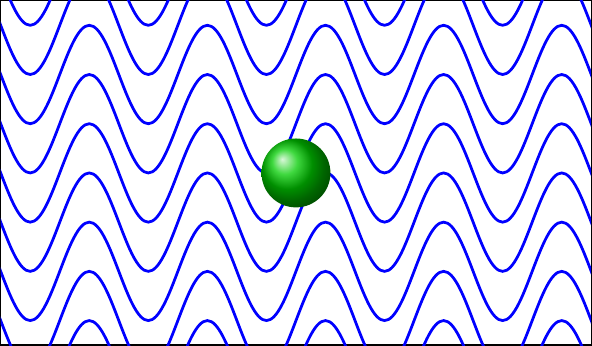}}	&%
\parbox[c]{\wavewidth}{\includegraphics[width= \wavewidth]{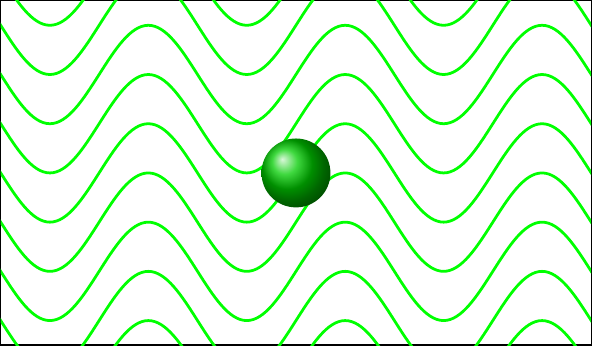}}	&%
\parbox[c]{\wavewidth}{\includegraphics[width= \wavewidth]{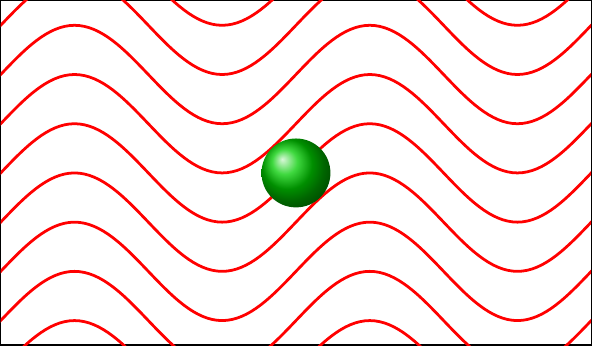}}		&%
$\Longleftrightarrow$	
&
\parbox[c]{\wavewidth}{\includegraphics[width= \wavewidth]{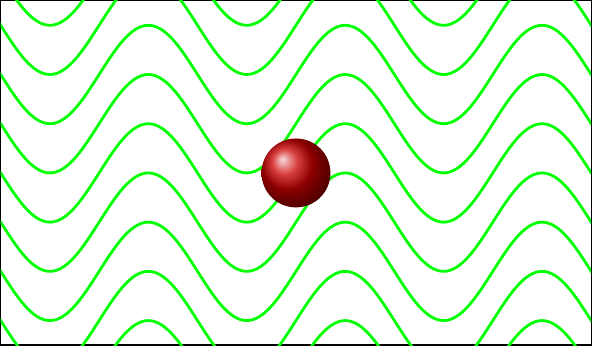}}	&%
\parbox[c]{\wavewidth}{\includegraphics[width= \wavewidth]{greenongreen}}	&%
\parbox[c]{\wavewidth}{\includegraphics[width= \wavewidth]{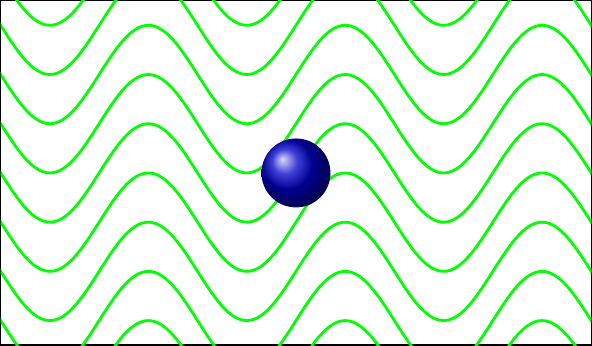}}
\\[1.5em]
\includegraphics[width= \levelwidth]{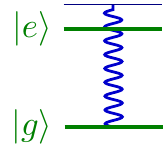}	&%
\includegraphics[width= \levelwidth]{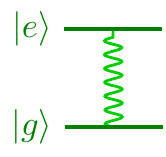}	&%
\includegraphics[width= \levelwidth]{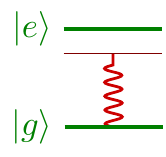}		&%
&
\includegraphics[width= \levelwidth]{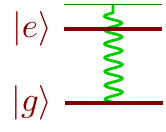}	&%
\includegraphics[width= \levelwidth]{greenongreen_levels}	&%
\includegraphics[width= \levelwidth]{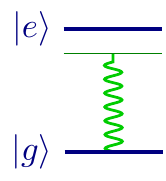}
\\[1ex]
\multicolumn{3}{c}{\scriptsize(a)~Field~picture} & &  \multicolumn{3}{c}{\scriptsize(b)~Detector~picture} \\
\end{tabular}
\end{center}
\caption{\label{fig:fielddetectpics}Schematic representation of the field picture and detector picture for a detector in an expanding universe exhibiting conformal invariance.  In the field picture~(a), expansion of the universe (shown left to right) causes the field modes to redshift, while the detector resonant frequency remains unchanged.  The field modes are defined with respect to cosmic time~$t$, using the metric from Eq.~\eqref{eq:FLRWmetriccosmic}, and the detector Hamiltonian~$H_0$ is independent of~$t$, satisfying Eq.~\eqref{eq:Schrodingert}.  Conformal invariance allows an equivalent description in the detector picture~(b), in which expansion of the universe (shown left to right) causes the detector resonant frequency to blueshift, while the field modes remain unchanged.  The field modes in this picture are defined with respect to conformal time~$\conft$, using the metric from Eq.~\eqref{eq:FLRWmetricconformal}, and the detector Hamiltonian~$H_0(\conft)$, given by Eq.~\eqref{eq:H0conft}, is now $\conft$-dependent, satisfying Eq.~\eqref{eq:Schrodingerconft}.  With additional modulation of the detector window function [Eq.~\eqref{eq:windowfunc}], both pictures result in exactly the same prediction for the physical response of the detector, as shown in Eq.~\eqref{eq:AFLRW}.}
\end{figure}

To summarize, the behavior of a detector with fixed energy gap~$E$, determined by~$H_{0}$, on a comoving trajectory in an FLRW universe in the the conformal vacuum may be simulated by a detector with a time-dependent energy gap, determined by $H_{0}(\conft)$, moving inertially in the Minkowski vacuum.  In addition, the detector's window function must also be modified: ${\wind(t) \mapsto F(\conft)}$.  The relevant equations for these substitutions are Eqs.~\eqref{eq:H0conft} and~\eqref{eq:windowfunc}, respectively.  This correspondence applies equally well to multi-detector effects, such as entanglement, and it may be exploited in the contruction of analog gravity experiments.

\section{Cosmology in an ion trap}
\label{sec:cosmologyions}

With the power of the detector picture, a vast number of gravitational models become experimentally accessible simply by modulating the detector parameters appropriately.  Our analog gravitational model of choice is the linear ion trap~\cite{James1998}.  Previous authors have considered ion traps for simulating de~Sitter spacetime~\cite{Alsing2005} (including the Gibbons-Hawking effect~\cite{Gibbons1977}) and cosmological particle creation due to a finite-duration period of expansion~\cite{Schutzhold2007}.  We will compare our proposal to these in due course.  We also note that trapped ions rotating around a ring trap have been proposed for detecting the acoustic analog of Hawking radiation~\cite{Horstmann2010}.  In what follows, we describe the ion-trap analogs of the field (in the Minkowski vacuum), an ordinary De~Witt monopole detector~\cite{DeWitt1979}, and a modulated detector.

We will go into some detail in describing the analogy between trapped ions and fields in spacetime in order to make certain nuances clear, which can have an important effect when generalizing the model to backgrounds other than Minkowski spacetime.  The reader is encouraged to notice, specifically, that the actual analog of the field~$\phi$ is a \emph{dimensionless} discrete field~$\phi_n$~[Eq.~\eqref{eq:phidef}], that the usual way of writing the ion-laser interaction~[Eq.~\eqref{eq:HIdef}] conflates interaction-picture time dependece~[Eq.~\eqref{eq:HIdeforig}] with additional time dependence from the laser, which has important consequences when trying to apply Eq.~\eqref{eq:H0conft}, that the Lamb-Dicke parameter~$\eta$~[Eq.~\eqref{eq:LDdef}] becomes time dependent when the trap frequency changes\protect\footnote{While this does not affect our proposal, it does affect Refs.~\cite{Alsing2005,Schutzhold2007} and any other field-picture proposals for ion-trap simulations of gravitational effects.} and when the laser frequency changes~[Eq.~\eqref{eq:kchange}], and that changing the laser frequency may, in some cases, also affect the Rabi frequency~$\Omega_0$, again through Eq.~\eqref{eq:kchange}.  We restore explicit use of~$\hbar$ in this section.

\subsection{Discrete phonon field}
\label{subsec:cosmologyions:normal}

In a linear ion trap, a string of up to about 10~ions is confined to one dimension using a combination of radio-frequency and static electric fields and capped at the ends by static electrodes~\cite{Leibfried2003}.  Linearizing the total potential created by the harmonic potential due to the trap electrodes and the mutual Coulomb repulsion between the ions, we can model the collective motion of the ions in a linear trap as a collection of coupled harmonic oscillators.  With a coupling matrix~$\mat A$ obtained from the linearized combination of these potentials, the Hamiltonian is given by~\cite{James1998}
\begin{align}
\label{eq:Hionlocal}
	H = \frac {1} {2 M} {\vec p^T \vec p} + \frac {M \nu^2} {2} \vec q^T \mat A \vec q \,,
\end{align}
where $\vec q = (q_1, \dotsc, q_N)^T$~is a column vector of operators corresponding to the displacement of each ion from its equilibrium position, $\vec p = (p_1, \dotsc, p_N)^T$~is a column vector of corresponding momenta, $M$~is the mass of each ion, and $\nu$~is the effective harmonic trap frequency provided by the trap electrodes (typically, $\nu \sim \text{a few~MHz}$~\cite{Leibfried2003}).  The coupling matrix~$\mat A$ is symmetric and positive-definite.  Thus, we can diagonalize it as~$\mat A = \mat B^T \mat \Lambda \mat B$, where $\mat \Lambda = \diag(\mu_1, \dotsc, \mu_N)$~is a diagonal matrix of positive eigenvalues, arranged in ascending order.  The orthogonal matrix~$\mat B$ defines the transformation to normal-mode coordinates~$\vec Q = \mat B \vec q$ and momenta~$\vec P = \mat B \vec p$.  Equation~\eqref{eq:Hionlocal} can be rewritten in these coordinates as
\begin{align}
	H 
	&= \sum_{p=1}^N \frac {P_p^2} {2 M} + \frac {M \nu_p^2} {2} Q_p^2 = \sum_{p=1}^N \hbar \nu_p\, a_p^\dag a_p\,,
\end{align}
where
\begin{align}
\label{eq:nup}
	\nu_p \coloneqq \nu \sqrt{\mu_p}
\end{align}
is the oscillation frequency of normal mode~$p$, and (since the normal modes oscillate independently) we have diagonalized this Hamiltonian using the standard prescription for a set of independent harmonic oscillators, ignoring the zero-point energy.  The normal-mode raising and lowering operators satisfy~$[a_p, a_{p'}^\dag] = \delta_{pp'}$, as is appropriate for independent bosonic modes.  The local oscillations~$q_m$ about the ions' equilibrium positions are given, in the interaction picture, by~\cite{James1998}
\begin{align}
\label{eq:qoftdef}
	q_m(\conft) = \sum_{p=1}^N \sqrt{ \frac {\hbar} {2 M \nu_p} } b_m^{(p)} (e^{-i \nu_p \conft} a_p + e^{i \nu_p \conft} a_p^\dag)\,,
\end{align}
where $m \in \{1, \dotsc, N\}$~labels the ion, and $b_m^{(p)}$ is an entry of~$\mat B$, which defines the spatial mode functions of the normal mode.\protect\footnote{Prefactor and phase conventions for~$q_m$ vary in the literature.}  We can define a unitless version of this local displacement oeprator, as well:
\begin{align}
\label{eq:phidef}
	\phi_m (\conft) \coloneqq \sum_{p=1}^N \frac {b_m^{(p)}} {{\mu_p}^{1/4}} ( e^{-i\nu_p \conft} a_p + e^{i\nu_p \conft} a_p^\dag )\,,
\end{align}
with $q_m(\conft) = \sqrt{ {\hbar} / {2 M \nu} } \phi_m(\conft)$.  This discrete field will serve as the analog of the Minkowski scalar field~$\phi(\vec x, \conft)$.  Notice that laboratory clocks tick off \emph{conformal time}, so the time-dependence of these modes is as it would be for modes in Minkowski spacetime, and the time coordinate is appropriately labeled by~$\conft$.  Because the coupling matrix~$\mat A$ only approximates a simple ``balls on a string'' coupling (characteristic of a bosonic field), the field will have a nonstandard dispersion relation, given by the frequencies~$\nu_p$ from Eq.~\eqref{eq:nup}, as a function of the normal-mode index~$p$.  This deviation from ideality means that the analogy with a massless scalar field is an imperfect one.

\subsection{Laser-induced coupling of vibrational and electronic states}
\label{subsec:cosmologyion:laser}

A feature of this model is that the internal electronic state of each ion can be turned into a local detector for the displacement of that ion.  To achieve this for a given ion, its internal state must become correlated with its linear displacement from equilibrium.  This correlation is provided by an external laser, which is used to drive an electronic transition between two meta-stable electronic levels~$\ket g$ and~$\ket e$, separated in energy by~$\hbar\omega_\atom$~\cite{Wallentowitz1996,Leibfried2003,James1998}.

The interaction between an external classical laser field and the $m$th~ion is described, in the dipole and rotating-wave approximation,\footnote{There are actually two different rotating-wave approximations that will be mentioned in this paper.  The first, discussed only here, is between the laser frequency and the electronic levels of the ion.  This one is always valid since the laser frequency~$\omega_\laser \sim \omega_\atom \gg \nu$.  References~\cite{Leibfried2003,James1998} discuss this in more detail.  The second rotating-wave approximation---which is what will be meant by all further instances of this term---is between the detuning in Eq.~\eqref{eq:Deltadef} and the normal-mode frequencies.  The validity of this approximation depends on these two frequencies, the coupling strength, and the time of the interaction.  See Refs.~\cite{Drummond1987,Zaheer1988} and citations therein for more details.} by the interaction-picture Hamiltonian~\cite{Leibfried2003,James1998}
\begin{align}
\label{eq:HIdeforig}
	H_I^{(m)}(\conft) &= -i\hbar\Omega_0 \left[\sigma_+^{(m)}(\conft)e^{+i[k\cos\theta q_m(\conft) - \omega_\laser \conft]}-\sigma_-^{(m)}(\conft)e^{-i[k\cos\theta q_m(\conft) - \omega_\laser \conft]}\right]\,,
\end{align}
where $\Omega_0$~is the Rabi frequency for the laser-atom interaction (typically, $\Omega_0 \sim 100~\text{kHz}$~\cite{Leibfried2003}), $k$~is the magnitude of the laser's wave vector~$\vec k$, which makes an angle~$\theta$ with the trap axis, $q_m(\conft)$~is the interaction-picture position operator for the $m$th~ion, and $\sigma_\pm^{(m)}(\conft) = e^{\pm i \omega_\atom \conft} \sigma_\pm^{(m)}$ are its interaction-picture electronic raising and lowering operators, where~$\sigma_+^{(m)}=\moutprod e g$ and~$\sigma_-^{(m)}=\moutprod g e$.  We may additionally absorb the laser's time-dependent phase~$e^{\mp i \omega_\laser \conft}$ into the interaction-picture operators~$\varsigma_\pm^{(m)}(\conft) \coloneqq e^{\mp i \omega_\laser \conft} \sigma_\pm^{(m)}(\conft) = e^{\pm i \Delta \conft} \sigma_\pm^{(m)}$, where
\begin{equation}
\label{eq:Deltadef}
	\Delta=\omega_\atom-\omega_\laser
\end{equation}
is the detuning of the laser below the atomic transition.  This allows us to write the interaction Hamiltonian as
\begin{align}
\label{eq:HIdef}
	H_I^{(m)}(\conft) &= -i\hbar\Omega_0 \left[\varsigma_+^{(m)}(\conft)e^{+ik\cos\theta q_m(\conft)}-\varsigma_-^{(m)}(\conft)e^{-ik\cos\theta q_m(\conft)}\right]\,,
\end{align}
which is the common form in the literature~\cite{Leibfried2003,James1998}.

The size of the rms~fluctuation in~$q_m$ as compared to the wavelength of the laser is measured by the \emph{Lamb-Dicke parameter}
\begin{align}
\label{eq:LDdef}
	\eta \coloneqq \sqrt{ \frac {\hbar k^2 \cos^2 \theta } {2 M \nu} } \sim \frac {\Delta x^{(m)}_{\text{rms}}} {\lambda_\laser} (2 \pi \cos \theta) \,,
\end{align}
where $\sqrt{\hbar/2 M\nu}$~is the ground-state rms~fluctuation of the center-of-mass mode of the ions.  This quantity is representative of the overall rms~fluctuations of the $m$th ion, denoted~$\Delta x^{(m)}_{\text{rms}}$, since the frequencies~$\nu_p$ for all higher normal modes of the ions remain within an order of magnitude of~$\nu$ for small numbers of ions (up to $ N \sim 10$), realistic for current experiments~\cite{James1998}.  Typical values of the Lamb-Dicke parameter are~\mbox{$\eta \sim$ 0.01~to~0.1}~\cite{Walls2008}.  When~$\eta \sqrt{\bar n + 1} \ll 1$, the so-called ``Lamb-Dicke limit'' ($\bar n$~is the average phonon number), the ion is well localized with respect to the wavelength of the laser, and we can expand the exponentials in Eq.~\eqref{eq:HIdef} to first order in~$\eta$, which is equivalent to first order in~$k \cos \theta\, q_m(\conft)$, giving
\begin{align}
\label{eq:HIexpand}
	H_I^{(m)}(\conft) &\simeq \underbrace{\hbar\Omega_0 \varsigma_y^{(m)}(\conft)}_{\text{carrier}} + \underbrace{\hbar\Omega_0 k\cos\theta\, q_m(\conft) \varsigma_x^{(m)}(\conft)}_{\text{sideband}}\,,
\end{align}
where~$\varsigma_x^{(m)}(\conft) = e^{+i \Delta \conft} \sigma_+^{(m)} + e^{- i \Delta \conft} \sigma_-^{(m)}$, and~$\varsigma_y^{(m)}(\conft) = -i e^{+i \Delta \conft} \sigma_+^{(m)} +i e^{- i \Delta \conft} \sigma_-^{(m)}$.  The first term corresponds to excitation of the transition directly by the laser, while the second couples the atomic transition to vibrational motion.

With the laser sufficiently detuned away from resonance ($\abs\Delta \gtrsim \nu$) and having a sufficiently narrow bandwidth (interaction time longer than $\nu^{-1}$), we can neglect the carrier term and keep only the sideband term:
\begin{align}
\label{eq:HIsideband}
	H_I^{(m)}(\conft) &\xrightarrow[\text{($\Delta \ne 0$)}]{\text{sideband}} \hbar\Omega_0 \eta \sum_{r=1}^N \frac {b_m^{(r)}} {\mu_r^{1/4}} (e^{-i \nu_r \conft} a_r + e^{i \nu_r \conft} a_r^\dag) ( e^{ i \Delta \conft} \sigma_+^{(m)} + e^{- i \Delta \conft} \sigma_-^{(m)})\,.
\end{align}
This Hamiltonian can be used to couple the atomic transition to the vibrational motion, with the detuning~$\Delta$ serving as the effective resonant frequency for the detector.  When $\Delta = \nu_p$ (called the first red sideband), this interaction models a detector that becomes excited when a phonon is absorbed---an ordinary phonon detector.  When $\Delta = -\nu_p$ (called the first blue sideband), it becomes excited when a phonon is \emph{emitted}---a rather unorthodox detector indeed.  After these interactions have completed, readout (projection) of the electronic state is accomplished using a bright laser that strongly couples the excited state to an auxiliary level, causing excited ions to fluoresce, while ions in the ground state remain dark~\cite{James1998}.

As long as the coupling strength~$\Omega_0 \eta$ is weak enough, no more than a single $\ket g \to \ket e$ transition will be excited for a given ion in the time during which the laser coupling is active.  This allows us to use perturbation theory to calculate the detector response function.  Strictly speaking, an infinite interaction time is needed for the simplified detector description given above to be exact; for finite times, the nonresonant terms (such as~$a_r^\dag \sigma_+$) can be nonnegligible, resulting in red-sideband detections even when the ions are in the collective ground state.  But this is what we want: this interaction has the form of a De~Witt monopole coupling~\cite{DeWitt1979}, and keeping these nonresonant terms is important for modeling the response of the modulated detector used to simulate the effects of expansion, the details of which we now discuss.

\subsection{Time-dependent laser coupling}
\label{subsec:cosmologyion:TDcoupling}

The usual construction, described above, produces a De~Witt monopole detector for the ions' vibrational quanta.  For our purposes, though, we need to introduce time dependence into our detector.  What we need, specifically, is a detector whose Hamiltonian is scaled by~$a[t(\conft)]$.  But we can't simply apply this directly to the laser interaction Hamiltonian because the effective energy gap of the ``detector'' is~$\hbar \Delta$, which is the difference between the atomic transition energy and the laser energy.  Therefore, what we need is to apply the following modulation:
\begin{align}
\label{eq:Deltaofconft}
	\Delta \to \Delta(\conft) \coloneqq a[t(\conft)] \Delta\,,
\end{align}
which, if we wish to vary only the laser frequency, can be achieved by
\begin{align}
\label{eq:omegaLofconft}
	\omega_\laser \to \omega_\laser(\conft) \coloneqq \omega_\atom - a[t(\conft)] \Delta\,.
\end{align}
Alternatively, one could choose to Stark shift the atomic frequency such that $\omega_\atom \to \omega_\atom(\conft) \coloneqq \omega_\laser + a[t(\conft)] \Delta$.  This may be more experimentally viable for some applications, but we will assume from now on that all modulation is done within the laser.

The only other change necessary to simulate an FLRW metric in the detector picture is a new window function.  Window functions correspond to modulation of the Rabi frequency~$\Omega_0$ of the atom, which is proportional to the amplitude of the interaction laser~\cite{James1998}.  Since a window function was needed anyway for initiating and ceasing the laser interaction, the requirement of a new window function presents no additional fundamental challenge.  These modifications result in the following interaction Hamiltonian:
\begin{align}
\label{eq:HIexpandmod}
	H_I^{(m)}(\conft) &\xrightarrow[\text{($\Delta \ne 0$)}]{\text{sideband}} \hbar\Omega_0 \eta F(\conft) \sum_{r=1}^N \frac {b_m^{(r)}} {\mu_r^{1/4}} (e^{-i \nu_r \conft} a_r + e^{i \nu_r \conft} a_r^\dag) ( e^{ i \Delta t(\conft)} \sigma_+^{(m)} + e^{- i \Delta t(\conft)} \sigma_-^{(m)})\,.
\end{align}
One other relevant thing to note is that the Lamb-Dicke parameter~$\eta$ depends on the laser wavenumber~$k$, which is proportional to~$\omega_\laser(\conft)$.  Thus, modulating the laser frequency will also modulate the coupling strength automatically through a time-dependent~$\eta(\conft)$:
\begin{align}
\label{eq:kchange}
	\frac {\eta(\conft)} {\eta} = \frac {k(\conft)} {k} = \frac {\omega_\laser(\conft)} {\omega_\laser} = 1 - \frac {\Delta} {\omega_\laser} \bigl( a[t(\conft)] - 1 \bigr)\,,
\end{align}
but since~$\omega_\laser \sim \omega_\atom \sim \text{1~GHz}$, while~$\Delta \sim \nu \sim \text{1~MHz}$, the fractional change in the coupling strength due to the frequency shift will be~$\sim 0.1\%$ unless the scale factor gets to be of the inverse of that same order.  This is unlikely, however, because such a large modulation (beyond a few tens of MHz) is both experimentally challenging~\cite{Salech1991,Yariv1989} and unnecessary because the detector would be so far detuned from the normal mode frequencies that it would (at best) detect nothing and (at worst) begin to excite extraneous atomic transitions.  Therefore, while the time-dependence of the Lamb-Dicke parameter must be considered in actual experiments, it can be easily compensated by suitably modulating the laser power.  We also note that a similar argument may apply to the Rabi frequency~$\Omega_0$, depending on whether the atomic transition being excited is dipole-allowed or dipole-forbidden; in the case of the latter, $\Omega_0$ also varies with the atomic transition frequency~\cite{James1998}, which is of the same order as the laser frequency, and thus the calculation in Eq.~\eqref{eq:kchange} applies to~$\Omega(\conft)$, as well.  Regardless of the experimental methods needed to achieve it, the new interaction Hamiltonian remains as in Eq.~\eqref{eq:HIexpandmod}.

At this point, we should take some time to talk about the approximations that could break down because of this additional modulation.  One of the biggest concerns is avoiding unwanted atomic transitions.  If the the bandwidth of the modulation, which is governed by the inverse of the detection time (width of the window function), is too large, this broadband response could overlap with unwanted transition energies.  In addition, the range over which the frequency varies could cause the same problem even if the modulation is slow and over a long period of time.  This issue strongly depends on the spectrum of the actual atoms used.  Typically~\cite{Leibfried2003}, however, $\nu \sim \text{1~MHz}$, while $\omega_\atom \sim \text{1~GHz}$, which should provide quite a large bandwidth over which the laser frequency may be swept.  Finally, the rate of modulation affects the feasibility of the scheme, as well.  Modulations up to~$\sim \text{MHz}$ can be achieved with an acousto-optic modulator, while faster rates (up to~$\sim \text{GHz}$) can be achieved using an electro-optic modulator~\cite{Salech1991,Yariv1989}.  We now describe several possible applications of our FLRW analog system.

\section{Experimental applications}
\label{sec:apps}

\subsection{Two distinct mechanisms by which expansion creates particles}

Birrell and Davies~\cite{Birrell1982} go to great lengths to distinguish between two effects that might be called ``particle creation'' due to expansion.  The first is the nonzero detection rate of an ordinary detector in what should be considered an ``empty'' universe when the detector is on \emph{during the expansion}.  If the expansion eventually stops, and the universe is asymptotically Minkowski, this effect is no longer seen.  This does not necessarily mean that particles are not detected in that case, just that this particular mechanism for particle production only applies \emph{during} expansion.

There is another type of ``particle creation'' that might be observed after a finite period of expansion has passed.  Imagine a universe starting asymptotically in the Minkowski vacuum for very early times.  Now imagine that expansion begins and ends smoothly, leaving the system in a nonequivalent vacuum but in a spacetime which is also asymptotically Minkowski for late times.  Particle creation occurs in this case because of the Bogoliubov transformation mixing annihilation and creation operators of the original vacuum into operators of a single type in the new vacuum ($a_\text{after} = \alpha a_\text{before} + \beta a_\text{before}^\dag$), thus ensuring that there will be real particles flying around after the expansion.

As we have formulated it, the detector picture can model the first case (detection during expansion) but not the second (detection after expansion).  The reason is the need for conformal invariance.\protect\footnote{Our work does not rule out a detector picture for cases that do not admit conformal invariance.  Our present formulation of it, however, requires this invariance.}  We have explicitly required that the vacuum we model in the detector picture be the conformal vacuum, which is (by definition) the ordinary Minkowski vacuum in conformal time.  If we were to model a finite-time expansion in the detector picture, particles would be detected during the expansion, but none would be detected afterward.\protect\footnote{To see this, just imagine a detector whose resonant frequency is first fixed at one value, then smoothly shifted another value and then fixed at that value, but only \emph{after} this process is the detector activated.  For long enough detection times (much longer than the inverse detector frequency), the rotating-wave approximation may be made, and no particles will be detected~\cite{Drummond1987,Zaheer1988}.}

An ion-trap analogy has been proposed for each of these effects.  Reference~\cite{Alsing2005} attempted to model the thermal response of a detector in de~Sitter spacetime (exponential expansion) by modulating the trap potential and thereby modifying the phonons' normal modes.  Since the proposal involved simulating the expansion by modifying the dynamics of the ions, the proposal uses the field picture.  There were several problems with the proposal, including incorrectly predicted dynamics, calculational mistakes, and ions that would fly out of the trap in a realistic experiment.  These problems were identified and discussed in Ref.~\cite{Schutzhold2007} (see endnote~[5] therein), which itself was an alternative proposal for witnessing particle creation in an ion trap---again by chirping the trap frequency---but this time the proposal was for the second type of particle creation in which real particles are created by the simulated expansion and detected after it has ceased.  This proposal does not suffer from the conceptual, calculational, and experimental problems of Ref.~\cite{Alsing2005}, but it also does not model the nonzero response of a detector \emph{during expansion}.

With the power of the detector picture, however, we can model a detector that is on during expansion---the original goal of Ref.~\cite{Alsing2005}---but with a completely static trap, instead embodying the entire effect of the expansion within the modulation of the laser used to couple the ions' vibrational motion to their electronic states.  In fact, we are not limited to de~Sitter spacetime in this picture; we can model \emph{any} scale factor as long as the required modulation is feasible experimentally.

\subsection{Single-detector response}

The detector response function~$A_m$ for the $m$th ion, to lowest nontrivial order in the Dyson series expansion, is given by
\begin{align}
\label{eq:Aionsinitial}
	A_m &= (\Omega_0 \eta)^2 \infint d\conft \infint d\conft' \, F(\conft) F(\conft') e^{-i\Delta[t(\conft) - t(\conft')]} \avg{ \phi_m (\conft) \phi_m(\conft') }\,.
\end{align}
This is the probability that an ion will get flipped to its excited electronic state by interaction with the field.  The interaction is assumed to be weak enough to only excite one transition from~$\ket g$ to~$\ket e$ during the detection, thus allowing the use of time-dependent perturbation theory~\cite{Menicucci2008a}.   We have extended the limits to~$\pm \infty$, with finite limits able to be enforced using the window function~$F(\conft)$, as required.  Also notice that the last term (in angled brackets) is just the ion analog of~$D^{+}_{\text{M}}(\conft, \conft')$.  Since the field is the ions' collective ground state, we can plug in Eq.~\eqref{eq:phidef} and evaluate
\begin{align}
\label{eq:DMioneval}
	\avg{\phi_m (\conft) \phi_m(\conft')} &= \sum_{p=1}^N \frac {\bigl[b_m^{(p)}\bigr]^2} {\sqrt{\mu_p}} e^{-i\nu_p ( \conft - \conft')} \,,
\end{align}
giving
\begin{align}
\label{eq:Aionsfieldeval}
	A_m &= (\Omega_0 \eta)^2 \sum_{p=1}^N \frac {\bigl[b_m^{(p)}\bigr]^2} {\sqrt{\mu_p}} \abs{ \infint d\conft \, F(\conft) e^{-i\Delta t(\conft)} e^{- i\nu_p \conft} }^2\,.
\end{align}
Notice that we can change integration variables back to~$t$ to obtain
\begin{align}
\label{eq:Aionsoft}
	A_m &= (\Omega_0 \eta)^2 \sum_{p=1}^N \frac {\bigl[b_m^{(p)}\bigr]^2} {\sqrt{\mu_p}} \abs{ \infint dt \, \wind(t) e^{-i\Delta t} e^{- i\nu_p \conft(t)} }^2\,.
\end{align}
This formula assumes $n=2$, which is the case for a (1+1)-dimensional spacetime---this being the most natural spacetime to simulate in a linear ion trap, which has one effective spatial dimension.  We note, however, that one-dimensional effects in spacetimes equipped with more than one spatial dimension can be simulated, as well, at the expense of additional powers of~$a(t)$ being included [see Eq.~\eqref{eq:windowfunc}].  The second form, Eq.~\eqref{eq:Aionsoft}, may be interpreted as the field-picture calculation of the same quantity as in the detector-picture formula, Eq.~\eqref{eq:Aionsfieldeval}, since in this integral the detector time dependence appears ordinary, being just~$e^{-i\Delta t}$, while the normal modes now experience time-dependent modulation, as~$e^{-i\nu_p \conft(t)}$.  Also notice that in the trivial case of no simulated expansion, $\conft = t$, and this quantity gives zero in the limit where $\wind(t) \to 1$ (unless $\Delta < 0$, indicating a blue-sideband detector), as a result of the rotating-wave approximation being applied to a detector in the vacuum.  We can, however, still achieve a nonzero response in this case for short enough detection times~\cite{Drummond1987,Zaheer1988}.

\subsection{De~Sitter spacetime and the Gibbons-Hawking Effect}

One of the most celebrated features of de~Sitter spacetime is the thermal response of a local detector to the conformal vacuum with a temperature proportional to the expansion rate:
\begin{align}
\label{eq:TGH}
	T_{\text{GH}} = \frac {\kappa} {2\pi}
\end{align}
in natural units.  This is the Gibbons-Hawking effect~\cite{Gibbons1977}.  It means that a single inertial detector in the conformal vacuum of de~Sitter spacetime will respond as if it were bathed in thermal radiation at a temperature~$T_{\text{GH}}$ in Minkowski spacetime.  An important physical difference between these two cases is that in the case of Gibbons-Hawking radiation, every inertial detector believes itself to be at rest with respect to the thermal bath it responds to---i.e.,\ no velocity-dependent Doppler shift is observed~\cite{Gibbons1977}---while a thermal state in Minkowski spacetime has a preferred rest frame, and detectors in relative motion with respect to this frame (even if still inertial) will perceive a Doppler-shifted spectrum~\cite{Birrell1982}.

It is this thermal response that the authors of Ref.~\cite{Alsing2005} wanted to simulate by chirping the trap potential and modifying the phonons' normal modes.  Since the proposal involved simulating the expansion by modifying the dynamics of the ions, it uses the field picture.  Later work corrected this proposal for a single ion~\cite{Menicucci2007a}, which again involved exponentially modulating the trap frequency.  The results were different from those calculated (incorrectly) in Ref.~\cite{Alsing2005} but still showed an effectively thermal response with a perceived thermal state with temperature~$T_{\text{GH}}$.\protect\footnote{This is measured by taking the ratio of the detector response functions at the first red and the first blue sideband, giving~$e^{-2\pi \Delta/\kappa}$, which is the desired ratio.}  Here we will instead propose to model the Gibbons-Hawking effect in the detector picture.

De~Sitter spacetime is an FLRW universe with an exponentially increasing scale factor in~$t$:
\begin{align}
\label{eq:aoftdS}
	a(t) = e^{\kappa t} \qquad &\Longleftrightarrow \qquad a[t(\conft)] = -\frac {1} {\kappa \conft}\,,
\intertext{with transformations to and from conformal time given, respectively, by}
\label{eq:conftdS}
	\conft(t) = -\frac {e^{-\kappa t}} {\kappa} \qquad &\Longleftrightarrow \qquad t(\conft) = -\frac {\ln(-\kappa \conft)} {\kappa}\,,
\intertext{which means that small intervals in each variable are given by}
\label{eq:conftdSderiv}
	d\conft = e^{-\kappa t} dt \qquad &\Longleftrightarrow \qquad dt = -\frac {d\conft} {\kappa \conft}\,,
\end{align}
where we have chosen~$t=0$ to correspond to~$\conft= -1/\kappa$.  (In addition, $t \to \infty$~corresponds to~$\conft \to 0^-$, and~$t \to -\infty$ corresponds to~$\conft \to -\infty$.)  As in all simulations we are proposing, the clock on the wall in the laboratory ticks off seconds of conformal time~$\conft$, while the simulated proper time~$\tau$ of the detector is equal to the simulated cosmic time~$t$ and given by Eq.~\eqref{eq:conftdS} in terms of~$\conft$.  Ideally we would like our detector to have an infinite detection time in~$t$ to get a perfectly thermal spectrum.  But this is not physical, so we will simulate it with a detection time that is sufficiently long.  Following the procedure proposed in Ref.~\cite{Menicucci2007a} (but modulating the detector instead of the trap frequency), we will model the physical detector as having a flat window function over a finite detection time from~$0$ to~$T$:
\begin{align}
\label{eq:windowdS}
	\wind(t) \simeq \Theta(t) - \Theta(t-T)\,,
\end{align}
where $\Theta(u) = 1$ for positive~$u$ and 0~for negative~$u$.  The equality is not strict because  we want smooth turn-on and turn-off for the detector.  The details of these functions is not important because the approximations we use allow us to model this detector as being on for a very long time:\footnote{We use the approximations from Ref.~\protect\cite{Menicucci2007a} rather than those of Ref.~\protect\cite{Alsing2005} because those of the latter (which correspond to rapid expansion of the trap initially) do not actually allow the limits on the window function to be extended as the authors claimed.}
\begin{align}
\label{eq:kappaapprox}
	\nu \gg \kappa \quad \text{and} \quad \nu e^{-\kappa T} \ll \kappa\,.
\end{align}
A look at Eq.~\eqref{eq:Aionsoft} reveals that in the field picture, the time-dependence of the field modes in an FLRW universe satisfies~$e^{-i\nu_p \conft(t)}$.  These limits, respectively,  correspond to an expansion that begins in a regime where the rotating-wave approximation is valid~(at $t=0$) but extends into a regime where the field modes have effectively stopped oscillating~(at $t=T$), meaning that particles will be detected~\cite{Drummond1987,Zaheer1988}.  The first approximation lets the lower limit of the window function be extended from~$t=0$ to~$t\to -\infty$ because the detector will see nothing during~$t<0$ anyway.   The second allows the upper limit to be extended from~$t=T$ to~$t\to \infty$ because the detector, which is assumed to be tuned to a frequency~$\Delta \sim \nu$, is far off resonance from the mornal-mode frequencies~$\nu_p$, which are all within an order of magnitude of~$\nu$~\cite{James1998}.

In the detector picture, this corresponds to sweeping the trap frequency from~$\Delta$ (the actual detector frequency we want to simulate) at $\conft=-1/\kappa$ to~$\Delta e^{\kappa T}$ at~$\conft = -e^{-\kappa T}/\kappa$, with intermediate values determined by
\begin{align}
\label{eq:DeltadS}
	\Delta(\conft) &= - \frac {\Delta} {\kappa \conft}\,. 
\end{align}
This is a polynomial modulation in~$\conft$, while the proposals in Refs.~\cite{Menicucci2007a} and~\cite{Alsing2005} require exponential modulation of the trap frequency.  The window function in conformal time (assuming~$n=2$) is
\begin{align}
\label{eq:windconfdS}
	F(\conft) &= -\frac {1} {\kappa \conft}\,.
\end{align}
Plugging these into the integral of Eq.~\eqref{eq:Aionsfieldeval} gives
\begin{align}
\label{eq:thermalint0}
	\int_{-\infty}^0 d\conft \, F(\conft) e^{-i\Delta t(\conft)} e^{- i\nu_p \conft} &= \int_{-\infty}^0 d\conft \, (-\kappa \conft)^{\frac {i \Delta} {\kappa} - 1} e^{- i\nu_p \conft}\,.\end{align}
Substituting~$u=-\kappa \conft$, $du = -\kappa\, d\conft$, $\beta=\Delta/\kappa$, and $\alpha = \nu_p / \kappa$, along with appropriate convergence factors~($\alpha \to \alpha + i\epsilon$, $\beta \to \beta - i\epsilon$, with~$\epsilon \to 0^+$), gives
\begin{align}
\label{eq:thermalintsubs}
	\frac {1} {\kappa} \halfint du \, u^{i \beta - 1} e^{i\alpha u} &= \frac {\Gamma(i \beta)} {\kappa (-i \alpha)^{i \beta}} = \frac {\Gamma(i \beta)} {\kappa \alpha^{i \beta}} e^{-\pi \beta/2}\,.
\end{align}
Plugging in for the substituted variables and plugging the final result back into Eq.~\eqref{eq:Aionsfieldeval}, along with the identity~$\abs{\Gamma(i \beta)}^2 = \pi/ [\beta \sinh(\pi \beta) ]$, gives
\begin{align}
\label{eq:AdS}
	A_m(\Delta) &= \frac {(\Omega_0 \eta)^2} {\kappa \Delta} \frac {2\pi} {e^{2\pi \Delta/\kappa} - 1} \sum_{p=1}^N \frac {\bigl[b_m^{(p)}\bigr]^2} {\sqrt{\mu_p}} \,,
\end{align}
which is a Planck spectrum at the Gibbons-Hawking temperature.  This formula may be found in Ref.~\cite{Alsing2005}, but it does not correspond to the procedure that the authors proposed.  It is also different from the result obtained in Ref.~\cite{Menicucci2007a} for the case of a single ion.  This is expected, though, since modulation of the trap frequency creates single-mode squeezing that has no analog in de~Sitter inflation.

There are several experimental hurdles to be overcome.  First, we want~$\kappa$ to be low so that we can use the approximations in Eq.~\eqref{eq:kappaapprox}, but we don't want it to be so low that detection events are prohibitively rare~($< 1\%$).  Second, the laser modulation rates must be within feasible limits.  Third, the electronic structure of the ions must be such that extraneous transitions are not excited while modulating the laser.  We can address the first two problems briefly here while noting that any real application will require a more detailed analysis involving the atomic spectra.

Combining the approximations from Eq.~\eqref{eq:kappaapprox} gives~\mbox{$e^{-\kappa T} \ll 1$}, which means that \mbox{$e^{-\kappa T} \sim 0.05$}~or less.  This means that~$\kappa \conft(T) \sim 0.05 {\kappa \conft(0)}$, and so~$\conft(T) - \conft(0) \sim \kappa^{-1}$, meaning the detector has to be on for roughly the inverse of the expansion rate.  This makes the bandwidth of the modulated laser spectrum~$\sim \kappa \ll \nu$, so the normal mode frequencies are still well resolved, and we don't have to worry about broadband excitation of extraneous atomic transitions.  The detector energy gap~$\Delta(\conft)$ (and thus the laser frequency~$\omega_\laser$) must be modulated over a range~$(e^{\kappa T} - 1) \Delta \sim \text{20~MHz}$ for $\Delta \sim \nu \sim \text{1~MHz}$~\cite{Leibfried2003}.  The laser power similarly needs to increase by a factor of~$\sim 20$ over the course of the simulation.\protect\footnote{Simulating a higher-dimensional spacetime will change this requirement; see Eq.~\eqref{eq:windowfunc}.}

While all of the above is achievable with current technology~\cite{Salech1991,Yariv1989}, the detection probabilities are prohibitively low due to the exponential scaling with~$\Delta/\kappa$, which we have required to be large in Eq.~\eqref{eq:kappaapprox} (since~$\Delta \sim \nu$).  Nevertheless, having established the theoretical connection to Gibbons-Hawking radiation for always-on detectors in our model, higher detection probabilities can be achieved using window functions of shorter duration and\slash or by using larger values of~$\kappa$.  The resulting response function will no longer be a Planck spectrum but can be calculated numerically.  This situation still models the analogous de~Sitter detector---just one with a finite-duration window function.

There are two goals, then, for achieving an experimentally accessible simulation.  The first is to make the absolute excitation probability (for the red sideband) high enough to be detectable; the second is to have some sort of signature of thermality.  To boost the detection rate, we will have to violate the approximations from Eqs.~\eqref{eq:kappaapprox}.  This means the limits of the integral in  Eq.~\eqref{eq:thermalint0} will be finite:
\begin{align}
\label{eq:thermalintfinite}
	\int_{\conft_\init}^{\conft_\final} d\conft \, F(\conft) e^{-i\Delta t(\conft)} e^{- i\nu_p \conft} 
	= \frac {1} {\kappa} \int_{u_\final}^{u_\init} du \, u^{i \beta - 1} e^{i\alpha u} = \frac {\Gamma(i \beta,-i \alpha u_\final) - \Gamma(i \beta,-i \alpha u_\init)} {\kappa \alpha^{i \beta}} e^{-\pi \beta/2}\,.
\end{align}
where $u_{(\init,\final)} = -\kappa \conft_{(\init,\final)} = e^{-\kappa t_{(\init,\final)}}$, and~$\Gamma(z,b) = \int_b^\infty dx\, x^{z-1} e^{-x}$ is the upper incomplete gamma function, which is a built-in special function in many computer algebra packages.  We define the time interval in terms of the \emph{simulated} initial and final cosmic times---$t_\init$~and~$t_\final$, respectively---in order to emphasize that this deviation from a perfect Planck spectrum is not an experimental artifact but results rather from simulating a detector that is on only for a finite duration (specifically, from~$t_\init$ to~$t_\final$) in de~Sitter spacetime.  Defining the regularized upper incomplete gamma function~\mbox{$Q(z,b) \coloneqq \Gamma(z,b)/\Gamma(z)$}, we can see that Eq.~\eqref{eq:AdS} gets modified to
\begin{align}
\label{eq:AdSfinite}
	A_m(\Delta; t_\init,t_\final) &= \frac {(\Omega_0 \eta)^2} {\kappa \Delta} \frac {2\pi} {e^{2\pi \Delta/\kappa} - 1} R(\Delta; t_\init, t_\final)\,,
\end{align}
where
\begin{align}
\label{eq:Rdef}
	R(\Delta; t_\init,t_\final) \coloneqq \sum_{p=1}^N \frac {\bigl[b_m^{(p)}\bigr]^2} {\sqrt{\mu_p}} \abs{ Q \left(i \frac {\Delta} {\kappa},-i \frac {\nu_p} {\kappa} e^{-\kappa t_\final} \right) - Q \left(i \frac {\Delta} {\kappa},-i \frac {\nu_p} {\kappa} e^{-\kappa t_\init} \right) }^2\,,
\end{align}
which can be evaluated numerically.  Extending the limits to~$\pm \infty$ gives~$A_m(\Delta) = A_m(\Delta; -\infty, \infty)$, in agreement with Eq.~\eqref{eq:AdS}.

In the case of effectively infinite-duration detectors, we can take the ratio of the red- and blue-sideband detection probabilities and find the signature of thermality:~$A_m(\Delta) / A_m(-\Delta) = e^{-2\pi \Delta / \kappa}$~\cite{Turchette2000}.  For finite detection times that are shorter than required for the detector to thermalize with the perceived Gibbons-Hawking bath---which is the regime we are required to work in---this relation no longer holds.  However, a new ratio can be calculated from Eq.~\eqref{eq:AdSfinite}:
\begin{align}
\label{eq:signaturefinite}
	\frac {A_m(\Delta; t_\init, t_\final)} {A_m(-\Delta; t_\init, t_\final)} &= e^{-2\pi \Delta / \kappa} \frac {R(\Delta; t_\init, t_\final)} {R(-\Delta; t_\init, t_\final)}\;,
\end{align}
which depends on when the detector is switched on and off but once again agrees with the fully thermalized result in the limit of long detection time.

\subsection{Multiple detectors}

The ion-trap analogy using the detector picture carries over to the case of many simultaneous detectors without modification.  This is because all terms in the Dyson series are integrals involving the interaction Hamiltonian, and such integrals can always be recast from cosmic time to conformal time, resulting in a new interpretation as the response of Minkowski-spacetime detectors with time-dependent frequencies, as long as the window function is also appropriately modified.  This leads us to suggest the possibility of studying the two-point correlations of scalar-field modes in de~Sitter spacetime by measuring the correlation function for simultaneous electronic excitations in multiple ions~\cite{Menicucci2008a} under the detector modulation prescription proposed here.  Such correlations are believed, in inflationary cosmology, to be responsible for structure formation in the early universe~\cite{Bardeen1983,Bassett2006}.  Two-component Bose-Einstein condensates have served in recent work~\cite{Fischer2004} as analog systems for studying inflationary cosmology.  Because of the restriction to conformally invariant settings in the current proposal, our goals are comparatively modest.  By measuring correlation functions in the detection probability for excitations in multiple ions~\cite{Menicucci2008a} under a detector modulation corresponding to various scale factors, we can analyze the spectrum of fluctuations that should result from the corresponding cosmological predictions~\cite{Bardeen1983,Bassett2006}.

Independently of the cosmological implications, Ref.~\cite{Retzker2005} proposes a method of swapping ground-state entanglement of a chain of ions to the electronic states of the ions at the ends of the trap using sonically local interactions (i.e.,\ detection events that happen faster than the speed of sound can propagate across the trap).  This is the ion-trap analog experiment of the proposal in Ref.~\cite{Reznik2005} to use the Minkowski vacuum of a quantum field to generate entanglement between distant local detectors through entirely local interactions with the field.  Using the detector picture, we can extend the proposal of Ref.~\cite{Retzker2005}---as well as any other proposal that seeks to model only the Minkowski vacuum---to an equivalent proposal that instead models any FLRW universe in the conformal vacuum.  This is particularly exciting because it brings into the realm of possibility performing the analog of the experiment proposed in Ref.~\cite{VerSteeg2009} to distinguish heating of a Minkowski field from de~Sitter expansion through comparing whether entanglement is generated between two local detectors.  The detector interactions used in Ref.~\cite{VerSteeg2009} are impractical both in the actual vacuum of outer space and in an ion trap because the detection probabilities would be far too low.  However, in an attempt to implement this analog experiment, the nonstandard detection techniques of Ref.~\cite{Retzker2005}, which help boost the entanglement significantly over that achieved by the ordinary monopole detectors, may be applied to the two cases of (a)~modulated energy levels and window functions, corresponding to de~Sitter expansion, and (b)~a heated field with no detector modulation.  We leave the details of this proposal to future work, but we include it in the discussion to illustrate the wide-ranging applicability of the detector picture in ion-trap simulations of quantum fields.

\section{Conclusion}
\label{sec:conc}

We have demonstrated an alternative ``picture'' for analog QFT in curved spacetime, which encodes the effect of curvature in the detector rather than in the motion of the analog quantum field, and we have used this technique to propose new and more accessible analog experiments for curved-spacetime QFT effects.  These include the Gibbons-Hawking effect for a single detector in de~Sitter spacetime, plus the possibility of modeling structure formation in inflationary cosmology and the entanglement signature of de Sitter~spacetime.

A question which may be raised in the mind of the reader is the degree to which the ion trap/curved spacetime analogy continues to hold when time has been re-defined in such a way to make the trap resemble Minkowski spacetime.  As one example, ordinary Minkowski spacetime has no cosmological horizon---one of the most conceptually important ingredients of, for example, de~Sitter spacetime.  Careful consideration of points like this show that the analogy still holds and that the horizon is actually still present.  The horizon in the detector picture is related to the fact that evolution from~$-\infty$ to~$+\infty$ in de~Sitter cosmic time corresponds to an evolution from~$- \infty$ to~$0$ in the analog Minkowski time variable~$\conft$.  The cosmological horizon is thus re-interpreted as the backward light cone extending from the detector location at $\conft = 0$ in Minkowski spacetime, which equally well shields the detector from some of the field modes which are necessary to define the vacuum state.

In a more general sense, however, all analog experiments are, at their root, quantum simulations.  In a simulated system, there is no reason to expect that simulated time must evolve in the same way as ordinary time.  Anyone who has done numerical simulations of time-dependent systems can attest to this.  Going between the two pictures can be thought of as running exactly the same simulation program on two different machines---while the simulation may be run at different rates (depending on the machine), the end result is precisely the same output, and the degree to which the output is analogous to something in the ``real world'' is identical.

\acknowledgments

We thank Greg Ver Steeg for stimulating discussions and ideas, Dave Kielpinski for consultation on the experimental feasibility of our scheme, and Tim Ralph for a forum in which to present and discuss our ideas.  We also thank the Defence Science and Technology Organisation (DSTO) for their support, as well as The University of Queensland and The Australian Research Council Centre of Excellence for Quantum Computer Technology, whose hospitality toward N.C.M. was instrumental in bringing this work to fruition.  Research at the Perimeter Institute is supported by the Government of Canada through Industry Canada and by the Province of Ontario through the Ministry of Research \& Innovation.


\bibliographystyle{bibstyleNCM}
\bibliography{IonFLRWpaper}

\end{document}